# Automated error correction in superdense coding, with implementation on superconducting quantum computer

**Kumar Nilesh**[1]**, Piyush Joshi**[2]**, and Prasanta Panigrahi**[3]

[1]**Department of Mathematics and Statistics, Indian Institute of Science Education and Research, Kolkata**
[2]**Department of Physics, Indian Institute of Space Science and Technology, Kerala**
[3]**Department of Physical Sciences, Indian Institute of Science Education and Research, Kolkata**

**Abstract**

Construction of a fault-tolerant quantum computer remains a challenging problem due to unavoidable noise in quantum states and the fragility of quantum entanglement. However, most of the error-correcting codes increases the complexity of the algorithms, thereby decreasing any quantum advantage. Here we present a task-specific error-correction technique that provides a complete protection over a restricted set of quantum states. Specifically, we give an automated error correction in Superdense Coding algorithms utilizing n-qubit generalized Bell states. At its core, it is based on non-destructive discrimination method of Bell states involving measurements on ancilla qubits (phase and parity ancilla). The algorithm is shown to be distributable and can be distributed to any set of parties sharing orthogonal states. Automated refers to experimentally implementing the algorithm in a quantum computer by utilizing unitary operators with no measurements in between and thus without the need for outside intervention. We also experimentally realize our automated error correction technique for three different types of superdense coding algorithm on a 7-qubit superconducting IBM quantum computer and also on a 27-qubit quantum simulator in the presence of noise. Probability histograms are generated to show the high fidelity of our experimental results. Quantum state tomography is also carried out with the quantum computer to explicate the efficacy of our method.

## 1 Introduction

First introduced by Bennett and Wiesner [1], quantum superdense coding is a noteworthy application of quantum information theory in the field of communication. It utilizes shared quantum entanglement as a resource and makes it possible to send classical information at a rate twice that of quantum information transmitted through the channel. Transmission of one single qubit can therefore result in communicating two classical bits worth of information. However, two maximally entangled quantum bits are required to be shared among the sender and receiver prior to the start of the communication protocol.

This protocol has been realized in several systems [2, 3, 4], and an extension to use generalized bell states (GBS) with $n$ qubits rather than just the 2 qubit EPR pair has also been developed [5]. More recently, Superdense Coding (SDC) has found its use in secure direct quantum communication, which utilizes distinct quantum properties of qubits and provides a new approach to secure communication than quantum key distribution (QKD) protocols [6, 7, 8]. This newfound security can also be extended to multi-party scenario [9, 10] and are claimed to provide better rates than QKD schemes [11].

However, like any quantum communication scheme, the superdense coding is also prone to errors introduced during transmission [12, 13]. Further, the decoherence in the entangled states makes them susceptible to a partial loss of entanglement. These errors caused by quantum channel and decoherence could also lead to change in states along with loss of entanglement, which significantly reduces the quantum advantage we theoretically calculate for various quantum protocols [14, 15]. These errors in the quantum states can be classified into three primary classes, arbitrary phase-change, phase-flip, and bit-flip errors. Appropriate error correction needs to be performed so as to render them useful for their assigned tasks in a more extensive network. Various quantum error correction techniques have already been proposed, but most of them require the use of a large number of qubits, thus increasing the computational and creation complexity of the algorithms. In this paper, we show how a generalized superdense coding protocol can be shielded against arbitrary errors such as bit/phase flip or relative phase errors, which can be of huge benefit for secure direct communication and other cases where superdense coding is used [16, 17]. Rectifying such errors using the well-studied quantum error correction codes [18, 19] comes at a substantial cost of overhead because the detection of errors in the state is generally done through majority voting over the encoded qubits residing in an enlarged Hilbert space compared to the initial state. To get around these cost restrictions, we will employ what is known as task-specific error-correction techniques, which provide complete protection over a limited class of errors in a restricted set of quantum states. Interestingly, the generalized Bell states utilized for superdense coding fit this description as these states follow certain symmetries, which allow it to be protected from the set of bit/phase flip and arbitrary relative phase shift errors [20]. These are:

- A GBS state that undergoes phase flip, bit flip, or both will still only be an (albeit different) GBS state.

- An arbitrary phase change (not a flip) will change the GBS to a non-maximally entangled state, which can be brought back to the set with ease using a single ancilla qubit.

At its core, this specific error correction is based on the method of non-destructive discrimination of Bell states involving measurement on ancilla qubits [21, 22]. This idea can then be used to perform automated error correction, wherein the state information on the ancilla is used to recover the assigned state as and when any error arises due to phase or bit-flip errors. Further, this procedure can also be extended to Generalized Bell States (GBS) as well as to the maximally entangled d-dimensional qudit state [23]. We will also see in this paper how these automated error corrections can be used in generalized superdense coding using local operations and classical communications. Here, automatically means that no measurements are necessary for this error correction technique, unlike the syndrome measurements in ordinary error correcting codes. Akin to this idea, in the context of superdense coding where the maximally entangled GBS qubits are distributed among two parties, we introduce an automated error-correction technique to correct errors from the above-described set and show its effectiveness at rectifying these errors on real physical qubits using the recently open-sourced 7 qubit *ibmq_nairobi*, an IBM cloud quantum computer. The map of qubits arrangement and read out assignment error for *ibmq_nairobi* is provided in Fig. 1. Characteristics (average) of qubits for this quantum computer is given in Table 1 and Ref. [24] provide benchmarking of *ibmq_nairobi* quantum computer.

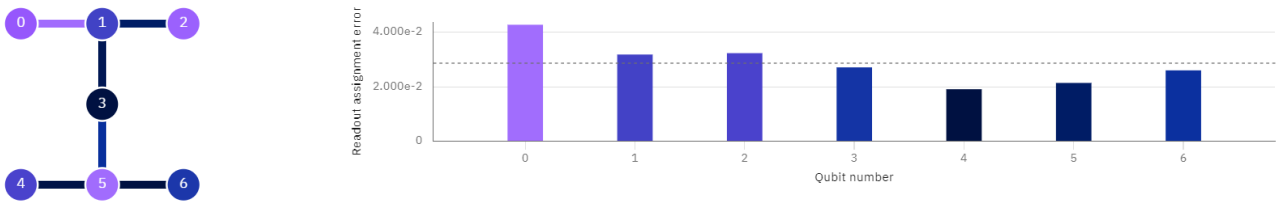

**Figure 1.** a. Arrangement of qubits in the 7-qubit superconducting *ibmq_nairobi* quantum computer. b. Readout assignment error for different qubits in *ibmq_nairobi*. It denotes the probability of a measurement returning an incorrect output.

| Qubit | Frequency | Anharmonicity | Readout length | ID error | CNOT error | Gate time |
|---|---|---|---|---|---|---|
| Q0 | 5.26 | -0.33983 | 5560.889 | 3.61E-04 | 0_1:2.969e-2 | 0_1:248.889 |
| Q1 | 5.17 | -0.34058 | 5560.889 | 4.08E-04 | 1_2:8.520e-3 | 1_2:426.667 |
| Q2 | 5.274 | -0.3389 | 5560.889 | 4.46E-04 | 2_1:8.520e-3 | 2_1:391.111 |
| Q3 | 5.027 | -0.34253 | 5560.889 | 5.29E-04 | 3_5:1.274e-2 | 3_5:277.333 |
| Q4 | 5.177 | -0.34059 | 5560.889 | 2.23E-04 | 4_5:6.443e-3 | 4_5:312.889 |
| Q5 | 5.293 | -0.34053 | 5560.889 | 3.75E-04 | 5_6:5.836e-3 | 5_6:341.333 |
| Q6 | 5.129 | -0.34044 | 5560.889 | 2.02E-04 | 6_5:5.836e-3 | 6_5:305.778 |

**Table 1.** Characteristics of qubit computers for *ibmq_nairobi*. Frequency and Anharmonicity are in *GHz* unit, and the unit for Readout length and Gate time is in *ns*.

The rest of the paper is organized as follows. Sec. 2 describes the generalized superdense coding and steps to perform automated error correction mathematically and builds a quantum circuit for its implementation. We then look at the traditional simple version of quantum superdense coding as a particular case in Sec. 3. In Sec. 4, we implement our automated error correction protocol in an actual 7 qubit superconducting quantum computer named *ibmq_nairobi*. The experimental results are then analyzed with the help of probability histogram and quantum state tomography. Finally, the paper is concluded in Sec. 5.

## 2 Generalized Version

A simple version of superdense coding involves two parties, in which one party wants to send two bits of classical information by just sending one quantum bit to the other party. Let's call the sender Alice and the receiver Bob. There is, however, a fundamental limitation to this type of protocol. We face with the restriction that the sender can only send even a bit of classical information by performing this protocol. The first step in this direction was provided by Saha and Panigrahi [25], who gave a scheme to send odd bits of classical information and later it was generalized to arbitrary bits of classical information by Dutta et al [26]. Here we present the most general form of super dense coding utilizing Generalized Bell States (GBS) [27], represented as:

$$\left|\psi_x^\pm\right\rangle = \frac{1}{\sqrt{2}}(|x\rangle \pm |\bar{x}\rangle)$$

where $x$ takes values from 0 to $2^{N-1}-1$ and $\bar{x} := 1^{\otimes N} \oplus x$ modulo 2. Using binary notation, this state can also be represented as

$$|\psi_N\rangle = \frac{1}{\sqrt{2}}\left(|0\rangle^{\otimes N} + |1\rangle^{\otimes N}\right)$$

This is a maximally entangled $N$-partite state. We will show how to perform generalized superdense coding with automated error correction in this section, and various special cases will be presented in the following sections.

As is the case with traditional superdense coding, both parties need to share a maximally entangled state. Therefore, along with the sender Alice and the receiver Bob we also consider a third party Charlie, whose task is to create a Generalized Bell State (GBS) and distribute the first $N-1$ qubits to Alice and the last qubit to Bob. The generation step requires a simple quantum circuit, where starting with a completely separable initial state ($|0\rangle^{\otimes N}$), Charlie applies Hadamard transformation followed by a sequence of CNOT gates.

Alice's goal is to send classical information by just sending her part of the entangled pair to Bob. But before she sends her qubit to Bob, she performs some local quantum operations on her qubits depending upon the classical message she wants to send to Bob. There is, however, no unique description of unitary operations for this encoding. A simplified encoding scheme is provided in Ref. [26]. Here the first qubit of Alice is used to encode the four different two bits of classical information, namely 00, 01, 10, 11 using the operation $\mathcal{O}_i = \{I, X, Z, iY\}$. Rest of the $N-2$ qubit of Alice is used to encode binary classical information using the operation $\mathcal{O}_k \in \{I \text{ or } Z, X \text{ or } iY\}$ where $k$ varies from 2 to $N-2$. Hence the total number of different possible unitary operations that Alice can perform on the $N-1$ qubits are $4 \times 2^{N-2} = 2^N$. Mathematically Alice's unitary operation on her side of qubits can be simply represented by the following operations,

$$\mathcal{O}_i \otimes \left(\bigotimes_{k=2}^{N-2} \mathcal{O}_k\right) \quad \text{where } \mathcal{O}_i \in \{I, X, Z, iY\}, \mathcal{O}_k \in \{I \text{ or } Z, X \text{ or } iY\}$$

It can be easily shown that these different unitary operations will generate $2^N$ different orthonormal states and hence forms the basis for this Hilbert space. Bob can therefore measure the states on this basis to decode the encoded message sent by Alice. Further, it is also shown that this is optimal in the sense that the Holevo bound of $H = \log_2 d = N$ is achieved in this case because for $N$-qubit GBS, we have dimension $d = 2^N$ [28, 29].

## 2.1 Automated error correction on generalized superdense coding

For states exhibiting certain symmetries, it is well known that these symmetries can be exploited for quantum error correction by designing symmetry-preserving, task-specific error correction circuits that give these states complete protection over random errors. To make our communication protocol error tolerant, we want to use the property of GBS being a restricted set of states with a restricted error class.

In the communication algorithm defined in the previous section, the orthogonality of the various generalized GBS states is one of the properties enabling dense communication between Alice and Bob. The orthogonality of these entangled states can also be used to discriminate them non-destructively, as shown in an experimentally verified method [21, 30]. This method uses ancilla qubits connected with our GBS qubits through gate operations and finally measured on a suitable basis to give complete phase and parity information about the state. Interestingly, it can also be shown that using these initially generated ancilla qubits, one can also *automatically* correct any bit/phase flip or arbitrary phase error occurring in the GBS states later in the communication channel without performing any measurement on the ancilla qubits, which is again experimentally verified [31]. This general automated error correction protocol is divided into three parts: the first part containing the partial entanglement discrimination of an $n$ qubit GBS using $n$ ancilla qubits (1 phase and $n-1$ parity ancillas) is performed by Alice. The second part is Bob completing the non-destructive phase and parity discrimination on the GBS state with possible errors. The third and final part is a three-step error-correction process, in which first, Bob corrects the arbitrary phase error (using a fresh ancilla) and then the phase flip, and finally bit flip error with the help of the ancilla qubits prepared after the discrimination stage.

As mentioned, SDC uses states with unique symmetry properties (or anti-symmetry in superposition) and can be discriminated non-destructively, allowing us to apply task-specific automated error correction to our transmitting qubits. But the difference in SDC from the general automated error correction of GBS is that the complete Bell State discrimination part cannot be performed locally by Alice alone because Bob holds the last qubit. However, we demonstrate in this paper that this discrimination algorithm has the advantage of being performed in a distributive manner [32]. This means that Alice can perform partial state discrimination on her side and send the incomplete information stored in ancillas, and

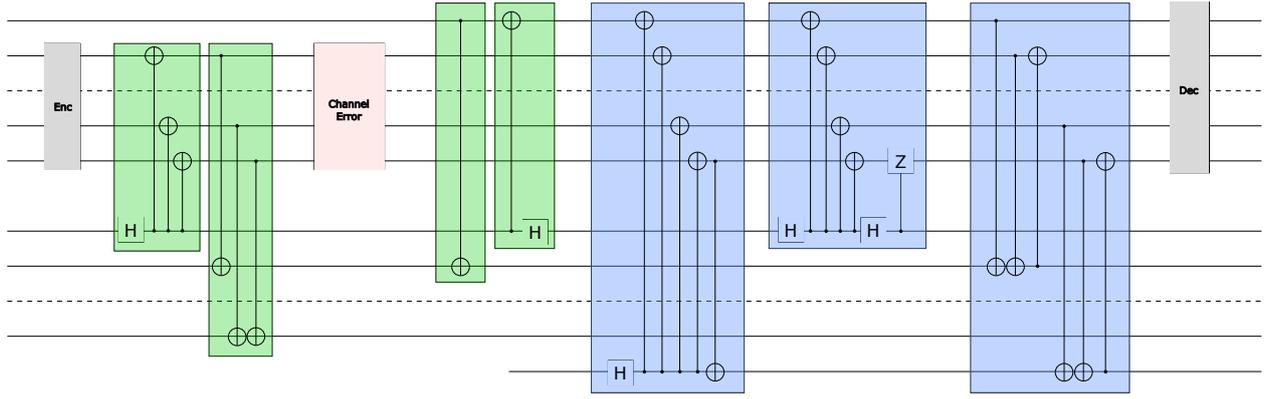

**Figure 2.** Automated error correction protocol for generalized bell states of *n* qubits distributed among two parties. Green state discrimination. Blue error correction. Define the indices of qubits and gates applied on ancillas.

Bob will perform a partial operation on his side. This provides the complete information, which is stored in the ancilla qubits and later these ancillas can be used to perform a complete Bell State discrimination and error correction. This advantage is possible because of the realistic assumption of errors occurring only in the communication channel and these particular qubits have already been used in state discrimination. This will be presented mathematically in the following subsection.

## 2.2 AEC SDC protocol

The complete automated error-corrected communication circuit is depicted in Fig. 2. We start with the operations that Alice performs on her end after encoding the information using the already known superdense coding protocol. For any GBS, the following two equations describe the partial Bell state discrimination done by Alice on the GBS in her possession in order to perform the error-correction protocol.

***Step 1-*** Alice performs partial phase discrimination on her qubits ($|\Phi\rangle_1$ to $|\Phi\rangle_{n-1}$) of the encoded $|\Phi\rangle$ using the ancilla qubit $|a_0\rangle$.

$$|\Phi_\phi\rangle_A \otimes |\phi\rangle \bigotimes_{k=2}^{k=N-1} |a\rangle_k = \left[\left(\bigotimes_{j=1}^{j=N-1} CX\left(|a\rangle_1, |\Phi\rangle_j\right)\right) \times \left(I_2^{\otimes N-1} \otimes H \otimes I_2^{\otimes N-1}\right)\right]\left(|\Phi\rangle_A \bigotimes_{k=1}^{k=N} |a\rangle_k\right)$$

This creates an entangled state $|\Phi_\phi\rangle$ of the encoded GBS with the phase ancilla $a_0$.

***Step 2-*** On the entangled state $|\Phi_\phi\rangle$, Alice then performs partial parity discrimination on $|\Phi_\phi\rangle_1$ to $|\Phi_\phi\rangle_{n-1}$ using $a_2$ to $a_N$ ancillas. Because Alice does not have access to the qubit $|\Phi_\phi\rangle_N$, we divide this step into two parts based on the resulting complete and incomplete parity ancilla qubits.

    **I.** The ancilla qubits $a_2$ to $a_{N-1}$ contains the complete parity information concerning their respective qubit pairs. Thus for $i = 1, \cdots, N-2$, we have:

$$|\Phi_\phi\rangle_A \otimes |\phi\rangle \bigotimes_{k=2}^{k=i-1} |a\rangle_k \otimes |p\rangle_i \bigotimes_{k=i+1}^{k=N} |a\rangle_k = \left[CX\left(|\Phi_\phi\rangle_i, |a\rangle_{i+1}\right) \times CX\left(|\Phi_\phi\rangle_{i+1}, |a\rangle_{i+1}\right)\right]\left(|\Phi_\phi\rangle_A |\phi\rangle \bigotimes_{k=2}^{k=N} |a\rangle_k\right)$$

    **II.** The parity information stored in the last parity ancilla $p_N$ will be incomplete as the last GBS qubit is with Bob.

$$|\Phi_\phi\rangle_A \otimes |\phi\rangle \bigotimes_{k=2}^{k=N} |p\rangle_k = \left[CX\left(|\Phi_\phi\rangle_{N-1}, |a_N\rangle\right)\right]\left(|\Phi_\phi\rangle_A \otimes |\phi\rangle \bigotimes_{k=2}^{k=N} |a\rangle_k\right)$$

**Communication error on the transmitted GBS qubits:** We consider arbitrary error $e$ on the $N-1$ GBS qubits when we are transferring $2N-1$ qubits ($|\Phi\rangle_1$ to $|\Phi\rangle_{N-1}$, $\phi$, and $p_1$ to $p_{N-1}$))

$$|\Phi_\phi\rangle \otimes |\phi\rangle \bigotimes_{k=2}^{k=N} |p\rangle_k \xrightarrow[Arbitrary\,Phase\,error]{Phase\,flip-Bit\,flip\,errors} |\Phi_\phi^e\rangle \otimes |\phi\rangle \bigotimes_{k=2}^{k=N} |p\rangle_k$$

Bob receives the error-affected GBS qubits along with partial parity and relative phase information in the form of the $N$ ancilla qubits. The discrimination part will be first completed by Bob before he performs the three-step error correction protocol.

***Step 3-*** Bob completes the phase and parity discrimination on $|\phi\rangle$ and $|p_N\rangle$ ancilla respectively using his $|\Phi_\phi\rangle_N$ qubit, hence separating these two ancillas from the previously entangled state. The resulting ancillas $\phi$ and $p_N$ contain complete phase and parity information, respectively, for the error-less initial encoded state of the GBS:

$$|\Phi^e\rangle \otimes |\phi\rangle \bigotimes_{k=2}^{k=N} |p\rangle_k = \left[ CX\left(|\phi\rangle, |\Phi_\phi^e\rangle_N\right) \times \left(I_2^{\otimes N} \otimes H \otimes I_2^{\otimes N-1}\right)\right] \left(|\Phi_\phi^e\rangle \otimes |\phi\rangle \bigotimes_{k=2}^{k=N} |p\rangle_k\right)$$

and

$$|\Phi_\phi^e\rangle \otimes |\phi\rangle \bigotimes_{k=2}^{k=N} |p\rangle_k = \left[ CX\left(|\Phi_\phi^e\rangle_N, |p\rangle_N\right)\right] \left(|\Phi_\phi^e\rangle \otimes |\phi\rangle \bigotimes_{k=2}^{k=N} |p\rangle_k\right)$$

We can now consider the erroneous GBS $|\Phi^e\rangle$ state separate from the discrimination qubits $\phi$ and $\bigotimes_{k=2}^{k=N} p_k$ and proceed to the error correction steps.

***Step 4-*** Utilizing a new ancilla $|a\rangle$, Bob rectifies arbitrary phase error in the $|\Phi^e\rangle$:

$$|\Phi^{e_1}\rangle |a'\rangle = \left[ (CX(|\Phi^e\rangle_N, |a\rangle)) \times \left(\bigotimes_{j=1}^{j=N} CX\left(|a\rangle, |\Phi^e\rangle_j\right)\right) \times \left(I_2^{\otimes N} \otimes H\right)\right] (|\Phi^e\rangle |a\rangle)$$

Any arbitrary relative phase the state might have acquired will be passed on to the ancilla $a'$. Note, however, that this step might introduce a phase flip error, which will be dealt with in the next step. The resulting state after this correction has error $e_1$, which can be a phase/bit-flip error.

***Step 5-*** Bob performs phase flip error correction on the $|\Phi^{e_1}\rangle$ state using the phase information stored in the ancilla qubit $\phi$:

$$|\Phi^{e_2}\rangle |\tilde{\phi}\rangle = \left[\left(I_2^{\otimes N} \otimes H\right) \times \left(\bigotimes_{j=1}^{j=N} CX\left(|\phi\rangle, |\Phi^{e_1}\rangle_j\right)\right) \times \left(I_2^{\otimes N} \otimes H\right) \times (CZ(|\phi\rangle, |\Phi^{e_1}\rangle_N))\right] (|\Phi^{e_1}\rangle |\phi\rangle)$$

The bit $\tilde{\phi}$ can then be discarded or can also be used to find out if there was a phase error or not. Note that $\tilde{\phi} = 1$ if there was any phase flip error, else $\tilde{\phi} = 0$. The resulting state $|\Phi^{e_3}\rangle$ now just contains bit flip errors.

***Step 6-*** Finally, to deal with the bit flip errors in $|\Phi^{e_2}\rangle$, Bob uses the $p_k$ ancilla qubits (where $i = 1, \cdots, N-1$) by applying the following operation:

$$|\Phi\rangle |\tilde{p}\rangle_i = \left[ (CX(|p\rangle_{i+1}, |\Phi^{e_2}\rangle_{i+1})) \times (CX(|\Phi^{e_2}\rangle_{i+1}, |p\rangle_{i+1})) \times (CX(|\Phi^{e_2}\rangle_i, |p\rangle_{i+1}))\right] (|\Phi^{e_2}\rangle |p\rangle_{i+1})$$

After performing all these steps of the protocol, we get the ideal encoded state $\Phi$ that Alice wanted to send to Bob, free of any errors acquired from the communication channel [33]. Bob can now easily decode this state to receive the classical information sent by Alice.

## 3 Automated error correction in two-qubit case

In this section, we focus on a simple two-qubit superdense coding and apply automated error correction as per the generalized protocol. The simplest version of superdense coding involves two parties, in which one party wants to send two bits of classical information by just sending one quantum bit to the other party [34, 35]. Similar to the generalized case both parties needs to share a maximally entangled state, here the EPR pair: $|\Phi^+\rangle = \frac{1}{\sqrt{2}}(|00\rangle + |11\rangle)$ [36, 37]. The local operations Alice needs to perform on her end based on the classical message to be sent are summarized in the Table 2.

| Alice's message | Quantum operation | ($\frac{1}{\sqrt{2}}$) Final state |
|---|---|---|
| 00 | $I$ | $|00\rangle + |11\rangle$ |
| 01 | $X$ | $|10\rangle + |01\rangle$ |
| 10 | $Z$ | $|00\rangle - |11\rangle$ |
| 11 | $ZX$ | $|01\rangle - |10\rangle$ |

**Table 2.** Single gate operations performed by Alice to encode the message.

The decoding operation performed on Bob's side is just the inverse of Bell state creation operations, i.e., Bob just apples a CNOT on Alice's qubit (with his qubit as control) followed by Hadamard operation on his qubit. Bellow, we will mathematically describe how our automated error correction can be integrated for this case to recover the encoded information reliably.

| Bell State | $\|p\rangle$ | $\|\phi\rangle$ |
|---|---|---|
| $\|\psi_1\rangle - \frac{1}{\sqrt{2}}(\|00\rangle + \|11\rangle)$ | $\|0\rangle$ | $\|0\rangle$ |
| $\|\psi_2\rangle - \frac{1}{\sqrt{2}}(\|00\rangle - \|11\rangle)$ | $\|0\rangle$ | $\|1\rangle$ |
| $\|\psi_3\rangle - \frac{1}{\sqrt{2}}(\|01\rangle + \|10\rangle)$ | $\|1\rangle$ | $\|0\rangle$ |
| $\|\psi_4\rangle - \frac{1}{\sqrt{2}}(\|01\rangle - \|10\rangle)$ | $\|1\rangle$ | $\|1\rangle$ |

**Table 3.** Summary corresponding parity and phase information stored in the ancilla.

We will start with deterministic and non-destructive Bell state discrimination and obtain the partial parity and phase information through which we will correct any arbitrary phase shift, phase flip, and bit flip error.

If $\Phi$ represents the EPR pair, then $\phi$ and $p$, its phase and parity information, respectively, can be obtained from the following two equations, which constitute the partial discrimination part:

$$|\Phi\rangle_A |\phi\rangle = \left[\left(CX\left(|a_1\rangle, |\Phi_j\rangle\right)\right) \times (I_2 \otimes H)\right](|\Phi\rangle_A |a_1\rangle)$$

and

$$|\Phi\rangle_A |p\rangle = \left[CX\left(|\Phi\rangle_1, |a_2\rangle\right)\right](|\Phi\rangle_A |a_2\rangle)$$

for $1 \leq i \leq N-1$.

Now coming to the error correcting part, Alice will send these ancilla bits to Bob. Bob will complete the discrimination part and proceed to the three-step error correction procedure.

$$|\Phi^e\rangle |\phi\rangle |p\rangle = \left[CX\left(|\phi\rangle, |\Phi^e_\phi\rangle_2\right) \times \left(I_2^{\otimes 2} \otimes H \otimes I_2\right)\right]\left(|\Phi^e_\phi\rangle |\phi\rangle |p\rangle\right)$$

and

$$|\Phi^e_\phi\rangle |\phi\rangle |p\rangle = \left[CX\left(|\Phi^e_\phi\rangle_2, |p\rangle\right)\right]\left(|\Phi^e_\phi\rangle |\phi\rangle |p\rangle\right)$$

Starting with arbitrary phase shift error, it will be corrected as follows:

$$|\Phi^{e1}\rangle |a'\rangle = \left[(CX(|\Phi^e\rangle_2, |a\rangle)) \times \left(\bigotimes_{j=1}^{j=2} CX\left(|a\rangle, |\Phi^e\rangle_j\right)\right) \times \left(I_2^{\otimes 2} \otimes H\right)\right](|\Phi^e\rangle |a\rangle)$$

where $a$ represents the ancilla. This step might now also introduce the phase flip error. And any phase flip error can be corrected by the following operations by utilizing $\psi$:

$$|\Phi^{e_2}\rangle |\tilde{\phi}\rangle = \left[\left(I_2^{\otimes 2} \otimes H\right) \times \left(\bigotimes_{j=1}^{j=2} CX\left(|\phi\rangle, |\Phi^{e_1}\rangle_j\right)\right) \times \left(I_2^{\otimes 2} \otimes H\right) \times (CZ(|\phi\rangle, |\Phi^{e_1}\rangle_2))\right](|\Phi^{e_1}\rangle |\phi\rangle)$$

Finally, to correct the bit flip error, Bob will use $|p_i\rangle$ and perform the following operations:

$$|\Phi\rangle |\tilde{p}\rangle = [(CX(|p\rangle, |\Phi^{e_2}\rangle_2)) \times (CX(|\Phi^{e_2}\rangle_2, |p\rangle)) \times (CX(|\Phi^{e_2}\rangle_1, |p\rangle))](|\Phi^{e_2}\rangle |p\rangle)$$

This completes the error correction part. Bob now can perform the decoding operation to decode the message sent by Alice through superdense coding algorithm:

$$[CX(|\Phi\rangle_2, |\Phi\rangle_1) \times (H \otimes I_2)](|\Phi\rangle)$$

## 4 Superdense coding protocol with automated error correction performed in IBM Seven-qubit quantum computer

In this section, we perform and analyze the experimental implementation of our automated error correction algorithm for superdense coding. For this, we use *imbq_nairobi* quantum computer, which is a 7-qubit superconducting quantum system [38, 39, 40] that can be accessed as an open source cloud computing system using Qiskit [41]. We perform out experiment a number of times for three different cases to check the efficiency of our protocol. For testing the proposed error correction algorithm, we only introduce the above-mentioned errors in the communication channel between Alice and Bob. Ideally, we should see a single peak in the probability histogram if it were not for the errors coming outside the communication channel in an actual quantum computer. But, as we are using a NISQ-era quantum computing device to check our communication algorithm, such devices have their own coherent, incoherent, and SPAM errors [42]. The effect of these errors should also be in mind while considering the result plots below. To obtain further information about the output results states of a given circuit, quantum state tomography is also done [43, 44, 45].

## 4.1 Error Correction in superdense coding with EPR pair

We implement out error correction protocol in superdense coding with EPR pair individually on different types of error and then perform error correction simultaneously in all three types of error in a single circuit implementation. Fig. 3 represents the circuit for arbitrary phase error correction. As defined in the previous section, the arbitrary phase correction operation is done by Bob alone by initiating an ancilla qubit locally and performing certain gate operations. Here we introduce a random phase shift error of $\pi/3$ in the communication channel.

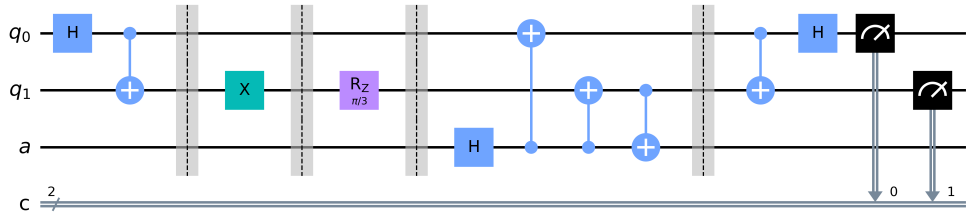

**Figure 3.** Implemented circuit for automated arbitrary phase shift error correction.

Phase flip error correction circuit is depicted in Fig. 4, where we introduced a single phase flip error, i.e., a $Z$ gate in the communication channel. As complete phase discrimination cannot be done locally by Alice, she completes her partial state discrimination and sends phase ancilla to Bob. Bob then completes the discrimination by the controlled-NOT and Hadamard gate. After the ancilla is in the required state, he then proceeds with the error correction. The state is then finally decoded to obtain the information.

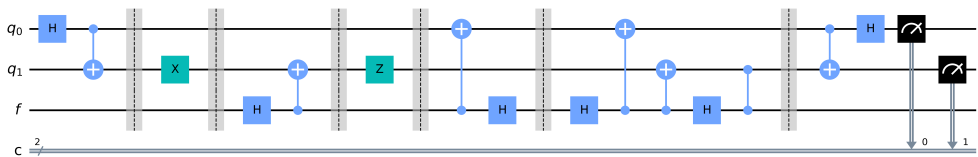

**Figure 4.** Implemented circuit for automated phase flip error correction using EPR pair.

Similar to the phase flip error, we introduce a single-bit flip error, i.e., $X$ gate in the communication channel. Here only one parity ancilla is required, and after encoding, Alice applies C-NOT from her control qubit to the target parity ancilla. This entangled ancilla then reaches Bob, where he completes the discrimination. Further, correction and decoding follow as presented in the Fig. 5.

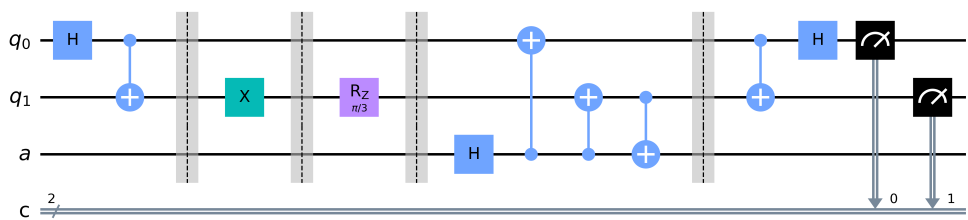

**Figure 5.** Implemented circuit for automated bit flip error correction using EPR pair.

We have also implemented, as shown in Fig. 6, our protocol in *ibmq_nairobi* where we correct all the above three types of error simultaneously in a single circuit.

## 4.2 Error Correction in superdense coding with 4-GHZ state

As an illustration of error correction in generalized superdense coding, we will look into the implementation of superdense coding algorithm with a 4 qubit GHZ state in the quantum computer [46]. An arbitrary phase shift error correction

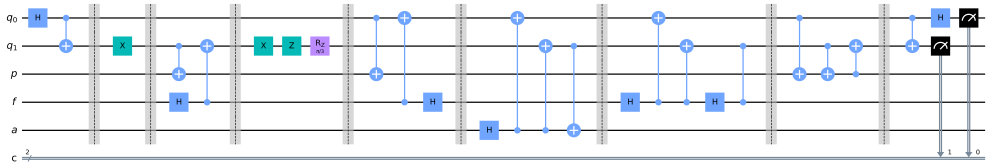

**Figure 6.** Implemented circuit for automated combined error correction using EPR pair.

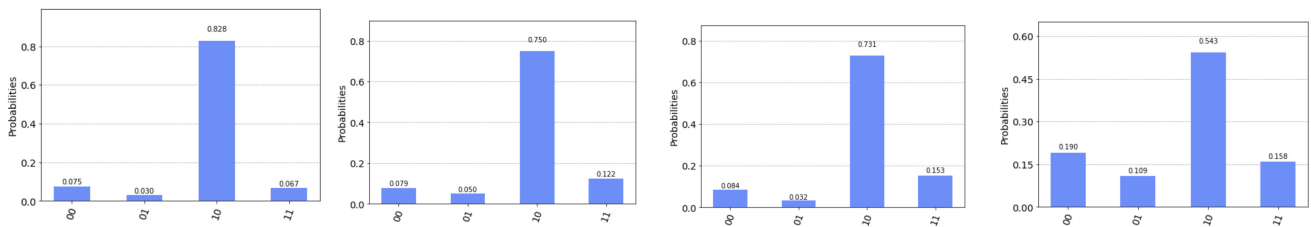

**Figure 7.** Experimentally obtained probability distribution after final measurement for superdense coding with EPR pair using *ibmq_nairobi*. This outcome agrees exactly with the theoretically expected value. a. Probability histogram for automated arbitrary phase error correction; b. Probability histogram for automated phase flip error correction; c. Probability histogram for automated bit flip error correction; and d. Probability histogram for combined automated error correction

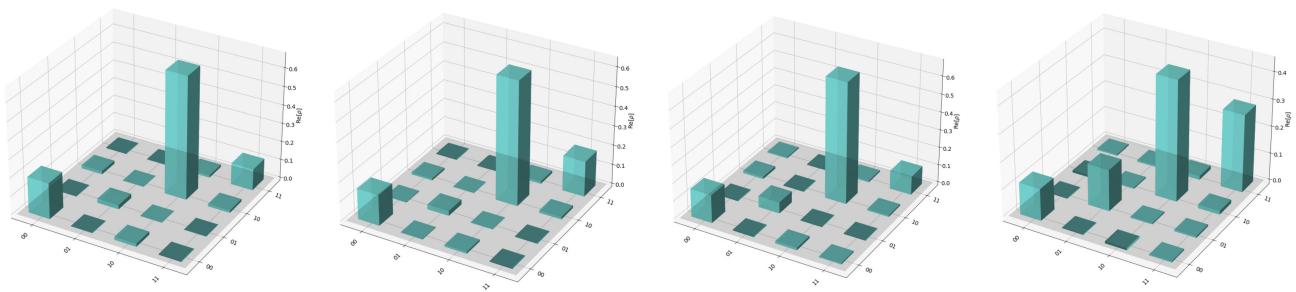

**Figure 8.** Experimentally obtained tomography for superdense coding with EPR pair using *ibmq_nairobi*. This outcome agrees exactly with the theoretically expected value. a. Experimental tomography for automated arbitrary phase error correction; b. Experimental tomography for automated phase flip error correction; c. Experimental tomography for automated bit flip error correction; and d. Experimental tomography for combined automated error correction

is done with the help of the circuit given in Fig. 9, and Fig. 10 represents the circuit for correction in phase shift error. Similarly, bit flip error correction for this algorithm is provided in the circuit Fig. 11. We have also implemented a combined error correction protocol for the 4-GHZ state algorithm, presented in Fig. 12.

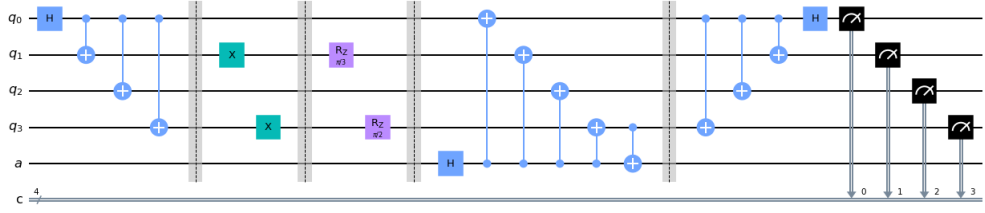

**Figure 9.** Implemented circuit for automated arbitrary phase error correction using 4GHZ state.

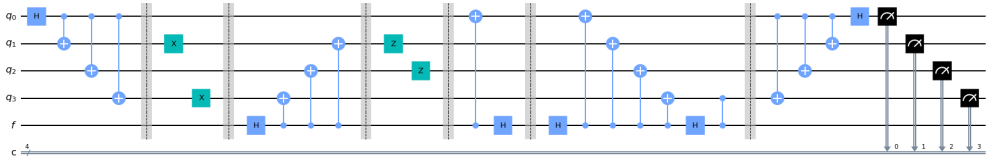

**Figure 10.** Implemented circuit for automated phase flip error correction using 4GHZ state.

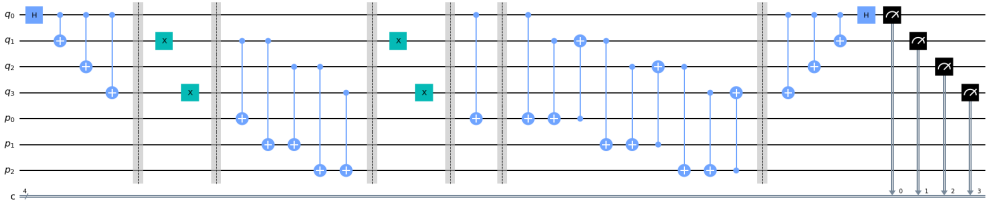

**Figure 11.** Implemented circuit for automated bit flip error correction using 4GHZ state.

### 4.3 Superdense Coding with GHZ-EPR pair

In this section, we will analyze a particular case of the generalized superdense coding, which involves a combination of various multipartite maximally entangled states. This is another way of removing the fundamental restriction of sending only even bits of classical messages through superdense coding. Practically, however, we can just discard the last bit to get an odd count, but this is not a very efficient way as Holevo bound is not reached in this case. Earlier, we have seen we can use GBS, but we can also use a combination of EPR pairs with GHZ state to send $2N + 1$ bits of a secret classical message by just using $N + 1$ qubits. In other words, they will share the following $N + 1$ qubits states:

$$|\Phi\rangle = |\psi_3\rangle \otimes |\phi\rangle^{\otimes N-1}$$

where $|\psi_3\rangle = \frac{1}{\sqrt{2}}(|000\rangle + |111\rangle)$ is a three-qubit GHZ state and $|\phi\rangle = \frac{1}{\sqrt{2}}(|00\rangle + |11\rangle)$ is an EPR pair. Initially, the last qubit of both the states is with Bob, while the rest is with Alice.

Let the classical binary string encoded by Alice be $b_{2N+1}b_{2N}\ldots b_2, b_1$. For this encoding, Alice performs the following local unitary operation at her end

$$|\Psi^k\rangle = \bigotimes_{i=1}^{N}(Z_{2i})^{a_i} \bigotimes_{j=1}^{N+1}(X_{2j-2})^{a_{j+N}} |\Phi\rangle.$$

Here $U_i$ denotes the operator $U$ acting on the $i$th qubit, and $U^0$ is considered to be Identity $I$ operator. It can be easily shown that the total number of encoded states $|\Psi^k\rangle$ with these operations are mutually orthonormal and forms the basis for $2^{2N+1}$ dimensional Hilbert space; hence this satisfies the superdense coding protocol and $k$ to varies from 0 to $2^{2N+1} - 1$. Bob can now measure the encoded states sent by Alice in this orthonormal basis to decode the binary classical string of length $2N + 1$.

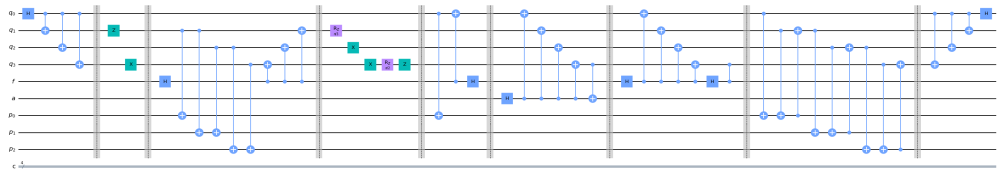

**Figure 12.** Implemented circuit for combined automated error correction using 4GHZ state.

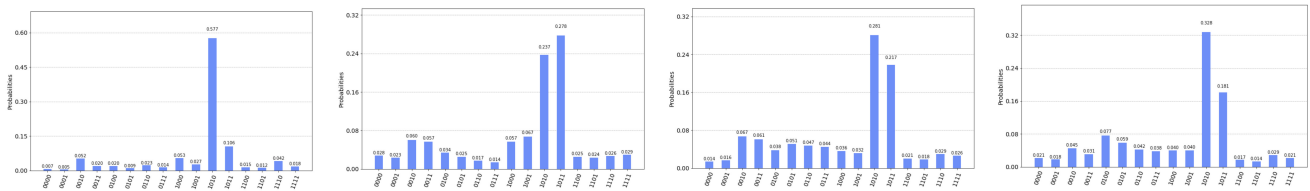

**Figure 13.** Experimentally obtained probability distribution after final measurement for superdense coding with 4GHZ state using *ibmq_nairobi*. This outcome agrees exactly with the theoretically expected value. a. Probability histogram for automated arbitrary phase error correction; b. Probability histogram for automated phase flip error correction; c. Probability histogram for automated bit flip error correction; and d. Probability histogram for combined automated error correction

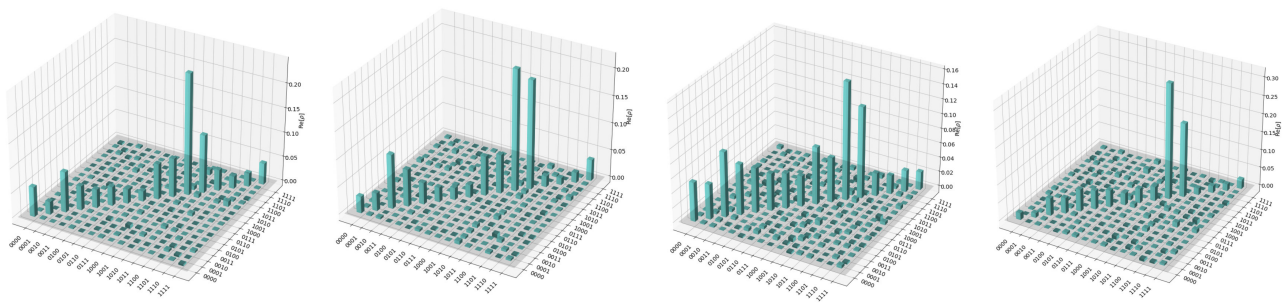

**Figure 14.** Experimentally obtained tomography for superdense coding with 4GHZ state using *ibmq_nairobi*. This outcome agrees exactly with the theoretically expected value. a. Experimental tomography for automated arbitrary phase error correction; b. Experimental tomography for automated phase flip error correction; c. Experimental tomography for automated bit flip error correction; and d. Experimental tomography for combined automated error correction

### 4.4 Error Correction in superdense coding with GHZ-EPR

As the state used for communication here has 5 qubits (EPR pair + GHZ state), for complete protection from the channel errors, it requires 5 more ancilla qubits (4 for parity and 1 for phase). The maximum number of open-source computing qubits is 7 as of writing this. Thus, we use a fake noise model simulator that simulates the noise of the currently working premium access quantum system, which has more qubits than the open-source systems [47]. We use the *FakeKolkata* noise simulating backend, which mimics the noise model of the 27-qubit *ibmq_kolkata* superconducting quantum system. Fig. 15 represents the circuit we implemented to perform our combined automated error correction for this GHZ-EPR pair superdense coding algorithm.

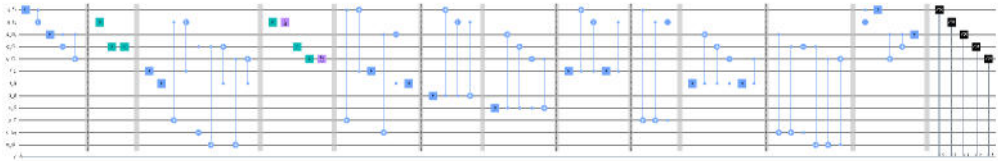

**Figure 15.** Implemented circuit for combined automated error correction using GHZ-EPR state.

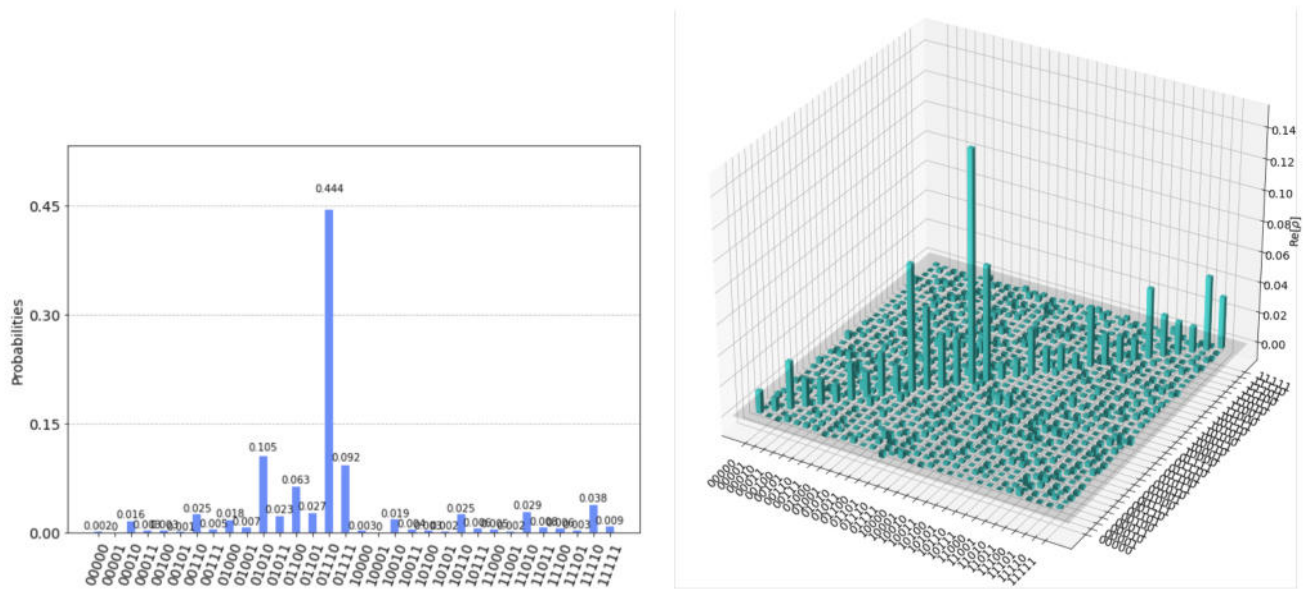

**Figure 16.** Experimentally obtained results for error correction in superdense coding with GHZ-EPR state using *ibmq_nairobi*. This outcome agrees exactly with the theoretically expected value. a. Probability histogram for combined automated error correction; and b. Experimental tomography for combined automated error correction.

## 5 Conclusion

Recent advancement in quantum computation has paved the way to replace classical computers completely and has found new applications in various fields, from cryptography to faster computation [48, 49, 50]. In this paper, we particularly focused on superdense coding, a very important application of quantum entanglement. But like any other quantum algorithm, they are prone to errors due to the delicate nature of qubits and entanglement between them [51]. Though various error correction codes are proposed, the requirement of a large number of qubits increases the computational and creation complexity of the algorithms, making them practically unusable with near-term quantum computers. This limitation motivates us to look for task-specific error-correction techniques that can provide complete protection over a limited class of errors in a restricted set of quantum states. This paper is one such example of these types of error correction techniques. The special symmetric properties of the states used in the superdense coding algorithm, more specifically the ability to non-destructively discriminate these states, allow us to use task-specific automated error correction techniques on this algorithm. We presented a step-by-step automated error correction protocol for the generalized superdense coding involving GBS. We primarily focused on three different kinds of errors, namely, arbitrary phase shift, phase flip, and bit-flip error, where we assumed errors to be independently distributed. We have also verified our results by implementing the protocol on a 7-qubit superconducting quantum computer named *imbq_nairobi*, which can be accessed as an open-source cloud computing system using Qiskit. We analyzed our experimental results using probability histogram and quantum state tomography, where we considered each type of error separately and also cases where all of them occurred simultaneously for three different scenarios. In one of these scenarios, we looked at the superdense coding algorithm that uses a mixture of different types of multipartite maximally entangled states.

Our paper also raises an important question that should we focus more on developing general quantum error-correcting codes, thereby increasing the complexity, or look for task-specific error-correcting codes that provide a significant advantage over the former? And for a more extensive algorithm, can we also implement these task-specific error correction techniques for each subroutine? Specifically for our protocol, it will be interesting to see if the same automated error correction can also be applied for quantum teleportation in some way, as it shares many similarities with superdense coding. And in general, for what other quantum algorithms can we exploit the symmetries and structures to generate these kinds of automated error correction schemes to gain a significant advantage in practical applications.